\documentstyle[12pt,titlepage]{article}

\newcommand{\PPPM}{P$^3$M}
\newcommand{\ie}{{\it i.e.}}
\newcommand{\cf}{{\it c.f.}}

\begin{document}


\title{A New Parallel N-body Gravity Solver: TPM }
\author{Guohong Xu \\
Princeton University Observatory, Princeton, NJ 08544-1001\\
email: xu@astro.princeton.edu}
\maketitle

\begin{abstract}
We have developed a gravity solver based on combining the well developed
Particle-Mesh (PM) method and TREE methods. It is designed for and has been
implemented on parallel computer architectures. The new code can deal with
tens of millions of particles on current computers, with the calculation
done on a parallel supercomputer or a group of workstations. Typically,
the spatial resolution is enhanced by more than a factor of 20 over the
pure PM code with mass resolution retained at nearly the PM level. This code
runs much faster than a pure TREE code with the same number of particles
and maintains almost the same resolution in high density regions.
Multiple time step integration has also
been implemented with the code, with second order time accuracy.

The performance of the code has been checked in several kinds of parallel
computer configuration, including IBM SP1, SGI Challenge and
a group of workstations, with the speedup of the parallel code
on a 32 processor IBM SP2 supercomputer nearly linear
(efficiency $\approx 80\%$) in the number of processors.
The computation/communication ratio is also very high ($\sim 50$), which
means the code spends $95\%$ of its CPU time in computation.

{\it Subject headings:}
cosmology: numerical --- N-body simulation ---
galaxies: clusters --- formation
\end{abstract}

\section{Introduction}

The performance of N-body simulations has become a very powerful tool to
investigate a wide range of astrophysical phenomena. Much effort has been
put into the search for efficient algorithms for N-body simulations.
The most direct approach, the particle-particle (PP) method is accurate,
but the CPU time required per time step scales as $\sim O(N^2)$, which becomes
very expensive when $N$ is more than a few thousand. Also, to some extent
the accuracy attained may be spurious if the system to be modeled is
expected to obey the collisionless Boltzmann equation
(\cf\ Hernquist and Ostriker 1992 for a discussion). In addition, N-body
systems
have proved to be very unstable (Goodman {\it et al} 1990).
For these reasons, efforts have been directed to the search for algorithms
to study self-gravitating systems using large number of particles
($\geq 10^7$), for which statistical properties are relatively stable,
and relaxation is minimal.

The particle-mesh (PM) method (Hockney \& Eastwood 1981, Efstathiou
{\it et al} 1985) uses the fast Fourier transform
(FFT) technique to solve Poisson's equation, thereby reduces the
operations per time step to $\sim O(M\log M)$, where $M$, the number of grid
cells, is typically approximately equal to the number of particles $N$.
With current computers, it is easy to simulate large number of
particles ($\sim 10^7$) using a PM code. Thus, good mass resolution
is obtained for most problems.  However, the spatial resolution is
constrained by the mesh spacing and usually is less than optimal.

The particle-particle-particle-mesh (\PPPM) method
(Eastwood \& Hockney 1974, Hockney \& Eastwood 1981,
Couchman 1991, Bertschinger \& Gelb 1991)
combines the advantages of both PP and PM methods.
In the \PPPM\ scheme, the force on each particle is the sum of a long range
force and a short range force.
The long range force is calculated by the PM method,
and the short range force as a correction is calculated by directly summing
the contribution from nearby neighbors.
Thus, the \PPPM\ method is much
faster than the pure PP method, and has a much higher dynamical range in
spatial resolution than pure PM method. However, \PPPM\ codes tend to be
slow in comparison with PM codes, since the bottleneck is in the short-range
force calculation. The short-range force calculation scales as
$N\langle N_B \rangle$, where $\langle N_B \rangle$ is the average
number of near neighbors, this makes the \PPPM\
method very slow when a system clusters and becomes non-linear.
Also, the requirements imposed by a fixed grid limit the
flexibility of this method, especially with regard to parallelization.

The TREE algorithm introduced by Barnes and Hut (1986) relies on a
hierarchical
subdivision of space into cubic cells. An octal tree is used to represent
the particle data. The root represents a cubic volume large enough to
contain all the particles in the system, this cell is subdivided into 8
cubic cells of equal volume. Each subvolume is subdivided into smaller
units, and this procedure is repeated until each cell at the lowest level in
the hierarchy contains either one or zero particles.
The force calculation in this
scheme is done by comparing the separation $d$ between the particle and the
cell center and the size of the cell $s$. If $\frac sd<\theta $,
where $\theta $ is
a fixed tolerance parameter, this cell is treated as one particle, and the
force between this cell and the particle is calculated by low order
expansion of the potential of the cell about its center-of-mass; otherwise
we go down one level if there is any, and perform the same comparison as
above between the children cells and the particle. This procedure is
repeated until it reaches the lowest level of the tree structure,
which contains only
zero or one particle. The tree structure that results from the above procedure
will have $\sim O(\log N)$ levels, and the force calculation for each
particle is proportional to $\log N$, rather than $N$, hence the CPU time
required to calculate the force on all particles scales as $N\log N$.
Since the overhead is high, the numerical coefficient make this
scheme considerably slower (factor $\sim 100$) than the PM scheme,
although both are nominally $O(N\log N)$.

In this paper, we propose and implement another hybrid method by taking
the advantages of PM method and TREE methods which we designate TPM.
The TPM approach is similar to the \PPPM\ method, in that the
short range force is handled by one method (TREE), and the long range
force is handled by another method (PM). Here, we treat particles in
overdense regions (primarily) as TREE
particles, and particles in low density regions as PM particles.
The forces on the PM particles are calculated by the PM method, and
the forces on TREE particles are the sum of an external force,
which is due to
the the particles outside the TREE and is calculated by PM method, and an
internal force, which is due to the particles
in the TREE containing the particle and is calculated by the TREE method.

Since the particles in high density regions have different dynamical time
scales from the particles in low density regions, it is necessary to
implement multiple time scale integration throughout
the whole simulation box. We implement multiple time scale algorithms
so that the TREE particles (in high density regions)
have much shorter time steps than PM particles (in low density regions)
with the time step optimized for each TREE.

\section{Our approach: TPM algorithm}

We have noted that the cost of two efficient algorithms
to solve the gravity equation for N-body problems, PM and TREE methods,
both scale as $N\log N$. PM codes have much higher
speed than TREE codes, but with limited spatial resolution, while TREE codes
have much higher spatial resolution, but with our current computers,
cannot easily do simulations with millions of particles, so for a
given mass density the mass per particle must be kept high,
\ie\ the mass resolution is poor.
We hope to achieve both high mass resolution and high spatial
resolution by developing a mixed
code of PM and TREE.  Gravity is a linear field, so forces calculated by
independent methods from different (overlapping) groups of particles can be
simply combined to determine the acceleration of test particles.

Discrimination between TREE particles in high density regions and PM
particles is made via a preset density threshold.
(details will be described in \S2.2).
Since the PM algorithm loses resolution when there are many
particles in one cell, this threshold should be set so that there are at most
a few PM particles in each cell.

In cosmological simulations, as many previous works have revealed,
an initially relatively uniform
field will develop filaments, pancakes and clumps (Peebles 1980).
Final structures develop into
fairly isolated regions (galaxies and clusters of galaxies),
thus we can remove those particles in a dense
region (cluster) from the PM simulation and simulate their evolution
with the high spatial resolution
TREE code. As there are many semi-independent structures
which exist in the simulated box, each
structure can be handled as a separate TREE, so they can be evolved in
parallel.
Based on this straightforward physical idea, we can write down the steps to
integrate particle motions as the following:

\begin{enumerate}
\item  assign all particles to grid with cloud-in-cell (``CIC'',
	Efstathiou {\it et al} 1985) scheme to calculate the PM density field.
\item  identify TREE particles as those above a certain density threshold.
\item  subdivide the particle space into a ``field'' of $N_{PM}$, PM
	particles and $M$ separate trees, each with $m_i$ particles
	($N=N_{PM}+\sum_{i=1}^Mm_i$).
\item  integrate the motion of the field PM particles using a large time
	step.  Force based on treating all particles as PM particles.
\item  integrate particle motion in each TREE separately
	(with many smaller time steps if necessary).
	Tidal force based on treating all particles not in a given
	TREE$_i$ as PM particles; local force from
	TREE algorithm for particles on a given tree.
\item  step time forward, go back to step 1.
\end{enumerate}

\subsection{Force decomposition}

As mentioned above, we divide the particles into PM particles and TREE
particles. Since the equation of gravity,
\begin{equation}
\nabla^2 \phi = 4\pi G \rho,
\end{equation}
is linear, we can decompose the force on a particle in a group of particles
into the sum of an internal force, which is due to the particles in
the same group, and an external force, which is due to the particles
outside this group, that is
\begin{equation}
{\bf F} = {\bf F}_{\rm internal} + {\bf F}_{\rm external}.
\end{equation}

When we calculate the force on a PM particle, we consider all the particles
to be in one group, and calculate the acceleration using the PM algorithm.
The total density $\rho_{\rm total}({\bf \vec n})$ on the grid is found using
CIC scheme. Then we obtain $\phi _{\rm total}({\bf \vec n})$ by solving
Poisson's equation using the FFT technique. The force on a PM particle is thus
\begin{equation}
{\bf F}_{\rm PM} = {\bf F}_{\rm total}^{\rm PM}
({\bf x}_i)=\sum\limits_{i,j,k}w_iw_jw_k\bigtriangledown \phi
_{\rm total}(i,j,k),
\end{equation}
where $w_i,w_j,w_k$ are weighing factors according to the CIC scheme
(e.g. Efstathiou {\it et al }1985).

When we calculate the force on a TREE particle, we consider the particles in
the TREE under consideration as a group, and all the other particles
(PM or TREE) as
another group. Thus the force on this TREE particle is the sum of
an external force and an internal force.
The internal force is calculated by the TREE method, and the
external force is calculated by PM method. When we calculate the external
force, we calculate the density of external particles by subtracting the
density of the particles in this TREE,%
\begin{equation}
\rho _{\rm external}=\rho _{\rm total}-\rho _{\rm internal},
\end{equation}
where $\rho _{\rm internal}$ is calculated in the same way as
we calculate $\rho _{\rm total}$ using the CIC scheme.
Then we get the external potential $\phi _{\rm external}$
by solving Poisson's equation using the FFT technique as we did
above, and then the external force ${\bf F}_{\rm external}$ using
the same scheme as above. Therefore, the force on a TREE particle is,
\begin{eqnarray}
{\bf F}_{\rm TREE} &=& {\bf F}_{\rm external}^{\rm PM}
+ {\bf F}_{\rm internal}^{\rm TREE} \nonumber \\
&=& \sum_{i,j,k} w_i w_j w_k \nabla \phi _{\rm external} (i,j,k)
+ {\bf F}_{\rm internal}^{\rm TREE}.
\end{eqnarray}

When we integrate the TREE particles using smaller time steps than
that of PM particles, the external force is updated every PM time step
rather than every TREE time step.  In detail, we know the external
(PM) force at the beginning and the end of the time interval, and so can use
interpolation to estimate the external force during the multi-step TREE
integration. This is reasonable because the
external force on these TREE particles does not change very much in one PM
step.

There are several reasons for doing this kind of force decomposition instead
of that of \PPPM. First, it is very easy to do parallelization among
clusters of particles by classifying particles into PM particles and TREE
particles. Secondly, the short range force calculation in the \PPPM\ algorithm
is not exact but is statistically correct (Efstathiou {\it et al} 1985), since
\PPPM\ needs a technique such as QPM (quiet particle mesh) to remove force
anisotropies which come from the long range force calculation in PM part.
In our TPM scheme, we do not have this problem and
the short range force is exact.
Thirdly, it becomes possible and easy to implement multiple time
scales, so that each cluster of particles can have its own time step. The force
calculation is just as straight-forward in this scheme as in the \PPPM\
scheme, but is different from that of \PPPM.

\subsection{TREE construction algorithm}

We construct TREEs in three steps: first, pick out TREE particles; second,
find isolated objects by examining the density field on the grid; third,
assign TREE particles to isolated objects, \ie\ individual trees TREE$_i$.
Each isolated object is treated as a separate TREE$_i$.

TREE particles are all in high density regions, but we may not want to pick
up all the particles in high density regions because this would leave
the density field outside with a sharp density hole. Thus we identify TREE
particles so that the density field left in the PM part is fairly uniform,
with its the maximum density lower than a certain threshold.

In sum, after calculating the density field in the PM part, we can
pick up particles
located in cells with density above a certain threshold level $n_{*}$,
so that the remaining PM density field satisfies the inequality
$n_{\rm PM} ({\bf x}) < n_*$.
Since the density calculation is made by accumulating particles in cells, we
define a particle to be TREE particle when it would make any of the nearby
eight
cells to which it would contribute density by CIC weighing have a density
greater than the defined threshold $n_{*}$. Alternatively, we could pick up
all the particles in a cell with density above the threshold density,
but this would, as noted above, leave a hole in the PM density field.

After identifying the TREE particles, we need an algorithm to group them into
separate subsets. The straight-forward method of grouping particles is by
finding a chain of nearest neighbor particles (friends-of-friends,
\cf\ Hockney and Eastwood 1981), but
this method is too slow and costly for our purposes. Since
we will in any case have calculated the density on the grid
for all the particles, and
the way we identify TREE particles ensure that these particles follow the
density distribution above certain threshold $n_*$, hence
we can construct TREEs using the density information on the grid.
We first estimate the local density peaks (on the grid)
as centers of a cluster of
objects. But we must invent criteria to decide
whether two or more local density peaks are to be included in the
same bigger cluster or not, \ie\ will they be treated in the same TREE?

Local density peaks are defined to be cells with density greater
than any of the eight
surrounding cells. However, a low density local density peak simply
means few particles around it, thus we will ignore those local density peaks.
After some numerical experiments, we find that we can set a density
threshold $\rho_{\rm peak}$, and we will only consider local density
peaks with density above $\rho_{\rm peak}$.
The algorithm adopted for defining $\rho_{\rm peak}$ is,
\begin{equation}
\rho_{\rm peak} = \min(\rho_{\max}, \bar\rho + \alpha \sigma_\rho),
\end{equation}
where $\rho_{\max}$ is the maximum density in the box,
$\bar\rho$ is the average density, $\sigma_\rho$ is the rms density
variation among all the cells. Here, $\alpha\approx 4.0$ in the
beginning of the simulation when $\sigma_\rho$ is small (and the density
field is close to Gaussian), and $\alpha \approx 10.0$ in the end of the
simulation when $\sigma_\rho$ is big.
The guideline to determine $\alpha$ is that we should not pick up
too many local density peaks, while allowing most of them to merge (the
criteria to merge local density peaks are discussed below).

After determining all the local density peaks,
we compute their overdensity radii by looking up the density distribution
in the surrounding cells. We set a certain density threshold
$\rho_{\rm radius}$ above which matter is
within the tidal radius of this object defined by the local density peak.
Then two objects would have strong interactions with each other if their tidal
shells touch each other, and they should be merged. This procedure guarantees
that the tidal forces, due to the external matter (treated via the PM
formalism), will never be very large, so that each tree is semi-independent of
its surroundings. Our numerical experiments find that the appropriate value
for $\rho_{\rm radius}$ is
\begin{equation}
\rho_{\rm radius} = \bar\rho + \beta \sigma_\rho,
\end{equation}
where $\beta \approx 1.0$.

Since it is not good to construct a TREE with too many particles, and
tidal forces can be handled fairly well by the PM code when two groups
are separately by 4-5 cells, we put restrictions on merging two groups
into one TREE, so that two groups with a separation of more than four cells
should not be merged, even if their tidal radii touch each other.

\subsection{Time Integration}

The equations of motion for particles in comoving coordinates are,
\begin{eqnarray}
\frac{dx}{dt} &=& v \\
\frac{dv}{dt} &=& -2H(t)v+a^{-3}F.
\end{eqnarray}
These equations can be integrated by the standard leap-frog scheme
(Efstathiou {\it et al} 1985),%
\begin{equation}
v(t+\Delta t)=\frac{1-H(t+\Delta t/2)\Delta t}{1+H(t+\Delta t/2)\Delta t}%
v(t)+\frac{a(t+\Delta t/2)^{-3}}{1+H(t+\Delta t/2)\Delta t}F(t+\Delta
t/2)\Delta t+O(\Delta t^3)
\end{equation}
\begin{equation}
x(t+3\Delta t/2)=x(x+\Delta t/2)+v(t+\Delta t)\Delta t+O(\Delta t^3).
\end{equation}

Time integration in the code is basically performed by the standard
leap-frog scheme. In addition, we allow for
different time steps between PM particles and TREE\ particles. PM particles
have a fixed (relatively large) time step through the simulation,
while TREE\ particles are allowed to
change time step from time to time. But to keep the second order accuracy,
we must be careful when the particles change time step. We found it very
convenient to keep the velocities of all particles synchronized at the end
of each PM step, so that second order time accuracy could be maintained.
That is,
at the beginning of each time step, PM particles have positions
$x(t+\Delta t/2)$ and velocities $v_i(t)$, while the TREE particles have
positions $x_i(t+\Delta t/2N_{\rm steps})$ and velocities $v_i(t)$.
Figure 1 gives an illustration how the time integration is done,
especially when the TREE time step changes.

When a PM particles becomes a TREE particle, its position must be
extrapolated from $t+\Delta t/2$ to $t+\Delta t^{\prime }/2$,
where $\Delta t $ is the PM time step,
and $\Delta t^{\prime} \equiv \Delta t/N_{\rm steps}$,
is the TREE time step.
This extrapolation can be kept to second order accuracy by the following
formula:%
\begin{eqnarray}
x\left( t+
\frac{\Delta t^{\prime }}2\right) =x\left( t+\frac{\Delta t}2\right)
+v(t)\left( \frac{\Delta t^{\prime }-\Delta t}2-2\frac{\dot a}a\frac{\Delta
t^{\prime 2}-\Delta t^2}8\right) \nonumber \\
+\frac{F(t+\Delta t^{\prime }/2)}{a^3}%
\left( \frac{\Delta t^{\prime 2}-\Delta t^2}8\right) +O(\Delta t^3).
\end{eqnarray}
Here, the force was calculated by PM method. When a TREE particle becomes a
PM particle, its position need to be extrapolated from
$t+\Delta t^{\prime }/2$ to $t+\Delta t/2$.
We do a similar extrapolation to the one given above. Notice that
in this extrapolation procedure, the force from the PM calculation is good
enough, because a particle will switch between PM particle and TREE particle
only when it moves to a region with marginal density near the threshold
density where a high force resolution is not essential.

When the TREE time step changes, we need to do the time integration of the
first TREE time step by the following formula, instead of leap-frog,
to keep second order accuracy:
\begin{eqnarray}
x\left( t+
\frac{3\Delta t^{\prime }}2\right) =x\left( t+\frac{\Delta t^{\prime \prime }%
}2\right) +v(t)\left( \frac{3\Delta t^{\prime }-\Delta t^{\prime \prime }}%
2-2 \frac{\dot a}a\frac{9\Delta t^{\prime 2}-\Delta t^{\prime \prime 2}}%
8\right) \nonumber \\
+\frac{F(t+\Delta t^{\prime }/2)}{a^3}\left( \frac{9\Delta
t^{\prime 2}-\Delta t^{\prime \prime 2}}8\right) +O(\Delta t^3),
\end{eqnarray}
where $\Delta t^{\prime \prime }$ is the old TREE time step, and
$\Delta t^{\prime }$ is the new TREE time step. Here, the force is calculated
by
the TREE method. The velocities are still integrated by the leap-frog scheme.
We
also restrict the TREE time step change so that $N_{steps}^{\prime }$ can
only change to $N_{steps}^{\prime } = N_{\rm steps}^{\prime\prime} \pm 1$
every time it changes. Remember that
$\Delta t'' \equiv \Delta t/N''_{\rm steps}$
and $\Delta t' \equiv \Delta t/N'_{\rm steps}$, the time step change is very
small when $N''_{\rm steps}$ is big. This helps to keep the time integration
accurate.

The time step for particles in one TREE can be estimated, again, with
the help of density field in the PM calculation. The average particle
separation in one cell is
\begin{equation}
\delta l_i\approx (m_i/\rho _i)^{1/3},
\end{equation}
where $m_i$ is the mass of the particle, and $\rho _i$ is the density of
the nearest grid point, and the average relative velocity of the particles
can be estimated by $\max (v_i,\sigma )$, where $\sigma $ is the rms
velocity of the TREE\ particles. So the time step is
\begin{equation}
\delta t_i=\beta \frac{\max (\delta l_i,\epsilon )}{\max (v_i,\sigma )},
\end{equation}
where $\epsilon $ is the softening length and $\beta \approx 0.1$. Then we
determine the TREE time step $\delta t_{TREE}\equiv \delta t_{PM}/N_{steps}$
so that 95\% of the TREE\ particles have $\delta t_i\geq \delta t_{TREE}$,
where $\delta t_{PM}$ is the PM time step. We could also take 100\%
confidence, but for the sake of keeping high time integration accuracy, the
time step should not change too much every time it changes. Since
$\delta t_i $ is a statistical quantity, if we
take the minimum of it as the TREE
time step, we are subject to big fluctuations. However, as $\beta $ does not
have to be a fixed value, we can change it so that $\max (\delta t_i)$ will
not change too much every time it changes, and we restrict $\beta <0.3$.
This is another scheme to determine the time step for TREE particles. We made
numerical simulations to test these two schemes, and the differences are
small.

Each TREE can have its own time step determined by the above method.
However, during our test simulation,
in order to maintain the highest time integration accuracy,
we used the same time step for all TREEs, \ie\ the smallest time step among all
the TREEs.
In fact, we did not save very much CPU time
by allowing TREEs to have individual time steps in our test runs.
In much larger runs there should be a significant gain.

\section{Parallelization}

Parallel programming is challenging. There are two
models in parallel programming, MIMD (Multiple Instruction
Multiple Data) and SPMD (Single Program Multiple Data). Since we have
essentially two different computations in our code, the code is
better suited to functional parallelism rather than to a simple data
parallelism, we choose MIMD as our programming model. Secondly, we must
decompose the data and decide which data is to be passed.
PVM (Parallel Virtual Machine) library is chosen as the message passing
library in the code, because it supports MIMD programming model and it
can be ported to many current systems. PVM is a well-accepted, public  domain
message-passing application developed at Oak Ridge National Laboratory.
Thirdly, the performance of the code is the most essential factor.
There are three major considerations for the performance of a parallel code.
(a) Scalability; here we refer scalability as the number of processors that
can be used without decreasing the relative efficiency, or in other
words, the megaflops/second on each processor remains nearly constant as you
increase
the number of processors.
(b) Computation/communication ratio; this is the ratio
of total communication time to total computation time, or the number of bytes
transferred to the number of operations on the data.
(c) Load balancing between the processors in use; some processors may
wait for some data to be ready, but we should minimize the waiting time and
the number of processors in waiting state.

We parallelize the TREE part of our code in two levels.
Since we have identified many physically isolated regions,
these regions can be evolved in parallel. Once we have created many
TREEs, we can distribute these TREEs among the processors so that the biggest
TREE is assigned to the fastest processor or to the processor with the lowest
load
when these processors are of same speeds, and the second biggest TREE is
assigned to the second fastest processor, until each free processor receives a
TREE. Once a processor finishes its assigned job, another remaining TREE is
handed over, until all the TREEs are processed.

Since the bottleneck of the computation is the TREE force calculation, this
level of parallelism results in very high computation/communication ratio.
But when the number of processors is bigger than or comparable to the number
of TREEs created, the load balance will be bad among processors, since the
small TREEs will be waiting for the biggest TREE to be finished.
This problem can be solved by the second level parallelism.

Most of the computation time $(\sim 97.5\%)$ in the TREE force
calculation is spent
in tree walk and force summation subroutines, while only about 2.5\% of CPU
time is spent in tree construction subroutines. Thus we can
parallelize the tree walk and force summation subroutines only and keep the
tree construction subroutine in serial. For one TREE, after tree construction,
this information is broadcasted to those processors which will process this
tree. Each processor calculates the forces for a fraction of the particles in
this tree, and the forces will be collected at an appropriate point.
This allows many processors to work efficiently on a single TREE.

Here is how we put the above two pieces of parallelization together.
First, distribute the TREEs to
different nodes; let them evolve the TREEs in parallel, once there is a free
node, either give it another TREE to process, if there is any,
or, if not, let it help other nodes to process a TREE
if there is any busy node, until all the TREEs are processed.
Figure 2 gives an illustration of how this parallelism works.
If the number of nodes exceeds the number of TREEs, those
nodes which do not get a TREE are treated as free nodes, and will help
those which have gotten a TREE. By doing this, we can potentially use thousands
of
nodes, however, the efficiency of PVM will decrease when hundreds of
workstations are grouped as a virtual machine, and we have to use more
efficient message passing tools available in those massively parallel
computers, such as MPL (Message Passing Library) in IBM SP1 and IBM SP2
machines. When we ran our code in the IBM SP1 machines, we replaced the PVM
subroutine calls with the corresponding MPL subroutines.

We parallelize the PM part of our code by decomposing the particle data among
all the processors, in addition, a parallel FFT subroutine is required.
Ferrell \& Bertschinger (1994) have developed the techniques to parallelize
the PM algorithm on the Connection Machine. We adopted their idea, and
made a naive implementation based on MPL subroutines. Although this
part suffers from a load balancing problem, it does not affect the speed
up very much,
since unbalanced load occurs if the density is concentrated in just a few
clusters, while the TREE part of the code tends to be slow also at this stage
of evolution. Thus TPM spends most of the CPU time on the TREE part when there
are big clusters in the box. However, the TREE part of the code does not
suffer from load balancing problem.

The speedup of the code with the number of processors was tested
using up to 32 nodes on the IBM SP2 machine at Maui High Performance
Computing Center (HPCC). The speedup curve is shown in Figure 3, and
we can see that we
maintain about $80\%$ efficiency up to 32 processors.

\section{Tests of the code}

The PM part of the code was developed as a standard particle-mesh code with
cloud-in-cell (CIC) density weighing method (Efstathiou {\it et al} 1985,
Cen 1992).
The TREE part of the code is the FORTRAN implementation of Barnes' tree
algorithm by Hernquist (1987). Periodic boundary conditions are achieved by
the Ewald summation method (Hernquist {\it et al} 1991,
Bouchet \& Hernquist 1988).
The actual code, kindly provided by Lars Hernquist,
is the TREESPH code (Hernquist and Katz 1989).

The first concerns for the code are the accuracy of force
calculation and the handling of the boundary condition.
Theoretically, the force can be
decomposed as we described in \S2.1; the actual code, however, may not
numerically keep the physics exact,
since the PM code and TREE codes have slightly
different Green's functions. To test the compatibility of the force calculation
from two codes, we put two particles in the simulation box randomly,
and calculated the force using the TREE code and PM code
separately. The PM code ($64^3$ cells) and the TREE code (with $\theta = 0.4$)
agree fairly well with each other
within $0.3 \%$ when the separation between the two particles
is above $\sim 4$ cell size (Figure 4).
The relative force error here is defined to be
\begin{equation}
{\rm Relative\ Error} = { |{\bf a}^{\rm PM} - {\bf a}^{\rm TREE}|
\over |{\bf a}^{\rm TREE}| }
\end{equation}
Our numerical experiments also shows
the optimal value for the tolerance parameter $\theta$
in the TREE code should be around $0.4$. If $\theta$ is too small, the
TREE code runs very slowly, while too big a $\theta$ produces too much
force error.
In the case of many particles, we have done the following test:
we put $N=1000$
particles randomly in the box, and calculated the forces on each particles by
the PM (cell number = $32^3$) and TREE ($\theta=0.4$) codes,
then calculated the
average relative force error between the forces from the two codes (Figure 5).
The results shows that the PM code allows surprisingly large force errors for
most of
the particles in the box; it becomes worse if there are more particles
in the box.

The threshold to identify TREE particles $n_\star$ is the most important
parameter in the TPM method. If we set $n_\star$ too low, this code behaves
almost
the same as a TREE code, which runs very slowly. If we set $n_\star$ too high,
this code is equivalent to a pure PM code, and we lose our resolution
and force accuracy.
The force error increases as $n_\star$ goes up. Thus, there is a ``best''
value such that the code has a reasonable resolution with fairly high speed.

In order to estimate the dependence of force error on the parameters,
several runs were performed for varying $N$, $n_\star$, $\theta$.
At each run, the accelerations computed through the TPM algorithm ($32^3$
cells, $\theta=0.4$) were
compared with those obtained by a pure TREE algorithm with $\theta=0.05$,
which is confirmed to be very close to those by direct summation
(Hernquist 1987). The force error was estimated for all the three components
separately. For each component of the acceleration, $a_i$, the mean error,
$\overline{\delta a_i}$, and the mean absolute deviation, $A(\delta a_i)$, are
defined by
\begin{eqnarray}
\overline{\delta a_i} &=& {1 \over N} \sum_{j=1}^{N}
\left( a_{i,j}^{\rm PMT} - a_{i,j}^{\rm TREE} \right), \\
A(\delta a_i) &=& {1 \over N} \sum_{j=1}^{N}
\left| a_{i,j}^{\rm PMT} - a_{i,j}^{\rm TREE} - \overline{\delta a_i} \right|,
\end{eqnarray}
and the absolute average acceleration, $\overline{a_i}$, is defined by
\begin{equation}
\overline{a_i} = {1 \over N} \sum_{j=1}^{N} \left| a_{i,j}^{\rm TREE} \right|.
\end{equation}
The ratio $A(\delta a_i)/\overline{a_i}$ as a function of $n_*$ for several
values
of $N$ is shown in Figure 6a for white noise model,
and in Figure 6b for CDM model at redshift zero ($N$ particles are randomly
picked up among $32^3$ particles from a simulation). For reasonably large
$N$ ($N \ge 10,000$) the typical error in the acceleration will be $\le 10\%$
for $n_*$ being a few,
far smaller in the TPM algorithm than in the standard PM approach (\cf\
Figure 5 or 6a with large $n_*$).

The tolerance parameter (Barnes et al 1981) is another important parameter in
this hybrid code. Typically we take a tolerance parameter
$\theta \approx 0.4 $ which leads to a force accuracy of
$\leq 0.2\%$ (Hernquist 1987) for the TREE particles.

A test run with $32^3$ particles in the standard CDM cosmology was performed
by using the TPM code with $32^3$ cells and $n_*=4.0$,
and the pure PM code with $256^3$
cells. The results at $z=0$ are rebinned to $32^3, 64^3$ and $128^3$ cells,
and the density
fields are compared on a cell-by-cell base for high density cells for
which $\min(\rho_1, \rho_2)/\bar\rho > 1.0\times(M^3/N)$, where
$M$ is the number of cells and $N$ is the number of particles.
This is shown in Figure 7.
First we note that the curves are reasonably symmetric and roughly Gaussian in
shape. The TPM code with cell number $32^3$ does not tend to
either overestimate or underestimate density fluctuations as
compared with a high resolution
($256^3$) PM simulation. Most of the difference between the two codes
shown in Figure 7 is due to displacement of small scale features between
the two simulations. The figure also shows that the TPM result tend
to have a
longer high density tail than PM$256^3$ result when rebinned to
finer grid.
After rebinning to $64^3$, where both codes should
be accurate, the difference between typical densities is still
approximately a factor of $2.5$.

\section{Resolution and Performances}

In order to estimate the resolution achieved by this code, we performed a
series of runs with the same initial conditions.
The initial condition was generated with a standard CDM model ($\Omega=1$,
$h=0.5$, $\Lambda=0$, $\sigma_8=1.0$), with $N=32^3$ particles in a box
of $L=50h^{-1}$Mpc.
We ran the TPM code with $n_* = 4.0, 8.0, 16.0, 32.0$, $32^3$ cells
in the PM part and $\theta = 0.3$ in the TREE part
(we call them TPM4, TPM8, TPM16 and TPM32 respectively hereafter).
In comparison,
we ran the pure PM code with $32^3, 64^3, 128^3, 256^3$ cells
(we call them PM32, PM64, PM128 and PM256 respectively hereafter).

First of all, we look at the global appearance of the whole box.
We project all particles to the $X-Y$ plane to get a view of the global
structure, they are shown in Figure 8 for PM runs and
in Figure 9 for TPM runs.
All the TPM results (with $n_*$ up to $32.0$) appear to be much better than
the PM64 result both in high density regions and low density regions.
The high density regions from the TPM runs with $n_*=4.0$ and $8.0$
appear to be better resolved than those from PM256.
For intermediate density regions, TPM (with $n_* \le 8.0$) resolves
the structure at least as good as high resolution PM (\ie\ PM256).

In order to compare quantitatively the resolution achieved by
various codes, we calculate the density field on $256^3$ grid points,
then we calculate the mass fraction
of particles locating at a grid above certain density level,
which is defined to be,
\begin{eqnarray}
f_{\rm mass} (\rho) &\equiv& \sum_{l=1}^{N} \sum_{\rho(i,j,k) > \rho}
W_{i,j,k} ( {\bf x}_l ) \\
&\equiv& \sum_{\rho(i,j,k) > \rho} \rho(i,j,k).
\end{eqnarray}
The results are shown in Figure 10 for all the runs.
The mass fraction function from TPM4 run is greater than
that from the PM256 run at high density end
(they intersect at a density of about $3.8 \bar\rho$),
which means the TPM4 run resolves clusters better than the PM256 run.
Even with $n_*=32.0$, the fraction mass curve of TPM run has a higher tail at
high density end than that of PM256 run, the rich clusters are also better
resolved in TPM32 run than in PM256 run (\cf\ Figure 8d and Figure 9d).
Since the PM code has a fixed ratio of resolution to grid scale,
the curves from the series of
PM runs (with $32^3, 64^3, 128^3, 256^3$ cells)
can be regarded as ``isochronic'' curves
indicating the resolution achieved by certain code. From this figure, we can
estimate the resolution we can achieve by choosing different values of
$n_*$ for the TPM code. The TPM algorithm has different resolutions for high
density regions and low density regions, actually $n_*$ determines the a
density $\rho_*$ above which we achieve high resolution. This density
$\rho_*$ can be estimated to the density at which PM and TPM results have the
same value of fraction mass. $\rho_*$ increases as $n_*$ slowly
when $n_*$ is small, and grows quickly when $n_*$ is big.

When we looked at the density contours of a cluster (Figure 11)
from the simulations, we can get a clear view of how well resolved
is the core of the cluster by the TPM code.
The method we make these density contours is the following:
cut a small cubic box with the object approximately at center
from the whole simulation box, and calculate the density field on the
$128^3$ grid points in the small box, project the density to a plane,
say $X-Y$ plane, and then get the density contours of this object.
We present in Figure 11 the results from TPM4 run and PM256 run.
Figure 11a (from TPM4) and 11b (from PM256) are the contours of a rich cluster,
and Figure 11c (from TPM4) and 11d (from PM256) are the contours of
a relatively faint cluster.
For both objects, the TPM4 results have more contour levels than the
PM256 results
indicating a higher central density for the object.
The contours are smoother in the TPM4 results than those in the PM256 results.
Inside the halo of the rich cluster, two objects are well resolved in the TPM
run while they are poorly resolved in the PM run (Figure 11a and 11b).

We will address the issue of the peculiar velocities of the clusters in
another paper. Here we just present the histogram of the peculiar velocities
of the particles at $z=0$ from several runs (Figure 12).
The velocity distribution from TPM4 is very similar to that
from PM256, both are apparently different from that from PM32,
especially at high velocity tails. This demonstrate that the TPM code traces
the velocity field with high precision.

The speed of this algorithm as a serial code is faster than a simple TREE
code, because we decompose the total number of particles into many groups,
$N = N_{\rm PM} + \sum_{k}^{M} N_{k}^{\rm T}$.
Actually, its speed is even comparable with the PM code with similar spatial
resolution (\cf\ Table 1).
Table 1 is compiled from a series of runs with identical
initial conditions on one processor of a Convex C3440.
Here, the CPU time per step for TPM
code is the ratio of total CPU time to the total number of
TREE steps. For this TPM run, $n_* = 8.0$, the total number of PM steps
is $200$ and the total number of TREE steps is $910$.
The spatial resolution for the PM code is estimated by $2.55/M$,
where $M$ is the number of cells. For TPM and TREE codes, the spatial
resolution is estimated to be twice of the softening length.

The parallelism of a code itself focusses on a key issue for
cosmological $N$-body simulation, \ie\ the wall clock run time for
one simulation.
We can estimate that a run with $256^3 \approx 1.68 \times 10^7$ particles
could be finished in 3-4 days using 64 IBM SP2 nodes.
As we can see from Figure 3, the code speeds up very nicely as the number
of processors increases.

\section{Conclusions and Discussions}

An ideal cosmological N-body algorithm should (a) have a wide dynamical range
in length and mass; (b) be capable of integrating the equations of motion
accurately and rapidly; (c) be able to efficiently utilize current
computer architectures. In this paper, we present our efforts to develop
a new TPM algorithm to approach this ideal. As we have demonstrated, the
TPM code retains the mass resolution of PM algorithm and the space resolution
of TREE algorithm, and it runs much faster than a pure TREE code.
The implementation of multiple time scale indicates that the new code should
integrate the equation of motion more accurately than other algorithms
for comparable computational effort.
The parallel TPM code speeds up very nicely in the parallel
supercomputers, and promises to run efficiently on massively parallel
systems.

The TPM algorithm will enable us to simulate cosmological phenomena
which requires high dynamical range, for example the formation and
structure of the cores of clusters of galaxies, gravitational lensing
in various cosmogonies. If we integrate this code with a high resolution
hydrodynamic code, we can simulate the galaxy formation with a scale
down to tens of kpc in a box of a hundred Mpc.
Although we developed the TPM algorithm in the
cosmological frame, it will be useful to study other astrophysical
phenomena. An obvious example is the interaction of galaxies.

The key parameter in the TPM algorithm is the density threshold parameter
$n_*$, it determines the ``depth'' of the structure we can resolve.
If we set variable density threshold for various high density regions,
\ie\ set a high $n_*$ for rich cluster and a low $n_*$ for faint cluster
in the same simulation, we might be able to resolve the
intermediate density region well
without necessarily spending a lot time in the highest density regions.
This might also help to resolve the filaments better while some high density
clumps are developing in the box.

It is a pleasure to thank my thesis advisor, J.P. Ostriker, for proposing
this scheme and for his inspiring advice and persistent guidance.
We thank L. Hernquist for providing us their
TREESPH code and some precious discussions.
We would also like to thank R. Cen and U. Pen for many useful discussions.
We acknowledge Cornell Theory Center for providing us the
opportunity to learn how to use SP1 machines.
Part of the simulations are done on the IBM SP2 machine in Maui HPCC
and Power Challenge machine in NCSA,
we would like to thank them for their support.
This work is supported by NSF grant AST91-08103 and
NSF HPCC grant ASC93-18185.

\newpage
\begin{center}
\begin{tabular}{c|ccccc}
\multicolumn{5}{c} {Table 1. Speed and Resolution Comparison} \\ \hline
\hline
   & cells & N & CPU/step & Length Resolution \\ \hline
TPM & $32^3$ & 32768 & 65.6 sec & 0.003 box size\\ \hline
PM & $256^3$ & 32768 & 84.5 sec & 0.010 box size\\ \hline
PM & $512^3$ & 32768 & 709 sec & 0.005 box size\\ \hline
TREE & -  & 32768  & 815 sec  & 0.003 box size\\ \hline
\hline
\end{tabular}
\end{center}

\newpage
\section*{Figure Captions}

\begin{description}
\item[Figure 1.] Illustration of time integration for the PM particles
	and TREE particles. The first PM time step contains 3 TREE
	time steps and the second PM time step contains 4. Notice
	that the positions are half time step behind the
	velocities.
\item[Figure 2.] Illustration of the parallel implementation
	of the TPM algorithm. Supposed there are 5 nodes in the
	parallel machine, and there are 8 TREEs identified at
	this time step. Solid box means the node gets the TREE
	and dashed box means the node is helping another node.
\item[Figure 3.] Speedup curve for the parallel TPM code running
	on IBM SP2 machine. The speedup is the ratio of wall clock
	run time on one processor to that on $N$ processors.
\item[Figure 4.] Force comparison between the PM code and
	the TREE code by putting two particles randomly in the box.
	TREE force are taken as the exact solution.
	Here, $64^3$ cells are used for the PM code,
	and $\theta = 0.3$ for the TREE code.
\item[Figure 5.] Force comparison between the PM code ($32^3$ cells)
	and the TREE code by putting 1000 particles randomly in the box.
	TREE force are taken as exact solution.
\item[Figure 6.] Force comparison between the TPM code ($32^3$ cells)
	with varying parameter $n_*$ and the pure TREE code.
	In (a) particles are put in the box randomly, while in (b)
	particles are randomly picked in a pool of $32768$ particles
	which is the result of a CDM run at redshift zero.
\item[Figure 7.] Distribution of $\log (\rho_2/\rho_1)$ where $\rho_1$ is
	the density field from PM $256^3$ run and $\rho_2$ is the density
	field from TPM $32^3$ ($n_*=4.0$) run. They are rebinned to
	$32^3, 64^3, 128^3$ cells separately.
\item[Figure 8.] Plots of two-dimensional projection of the final particle
	positions in a series of PM simulations:
	(a) PM$32^3$, (b) PM$64^3$, (c) PM$128^3$, (d) PM$256^3$.
\item[Figure 9.] Plots of two-dimensional projection of the final particle
	positions in a series of TPM$32^3$ simulations:
	(a) $n_*=4.0$, (b) $n_*=8.0$, (c) $n_*=16.0$, (d) $n_*=32.0$.
\item[Figure 10.] Mass fraction function from a series of runs:
	PM$32^3$, PM$64^3$, PM$128^3$, PM$256^3$, and
	TPM$32^3$ with $n_*=4.0, 8.0, 16.0, 32.0$. They are all rebinned
	to $256^3$ grid to calculate the density field.
	Here $\bar\rho = M^3/N$, with $M=256$ and $N=32768$.
\item[Figure 11.] Contours of images of a rich cluster and a faint cluster
	in CDM cosmology from the runs of PM$256^3$ and TPM$32^3$ with
	$n_*=4.0$.
	This figures are made such that (a) and (b), (c) and (d) have
	a small box at exactly the same position with same size respectively.
	(a) A rich cluster from TPM$32^3$ ($n_*=4.0$) run;
	(b) same cluster as (a) but from PM$256^3$ run;
	(c) a faint cluster from TPM$32^3$ ($n_*=4.0$) run;
	(d) same cluster as (c) but from PM$256^3$ run.
\item[Figure 12.] Peculiar velocity distributions from a series of runs:
	PM $32^3$, PM $256^3$ and TPM $32^3$ ($n_*=4.0$).
\end{description}


\end{document}